\documentclass{emulateapj}

\lefthead{Bekki \& Chiba}
\righthead{Globular cluster formation}

\begin{document}
\title{Massive stars and globular cluster formation}

\author{Kenji Bekki} 
\affil{
School of Physics, University of New South Wales, Sydney 2052, Australia}

\and 

\author{Masashi Chiba} 
\affil{
Astronomical Institute, 
Tohoku University, Sendai, 980-8578, Japan}

\begin{abstract}

We first present chemodynamical simulations
to investigate  how stellar winds of  
massive  stars influence early dynamical and
chemical evolution of forming
globular clusters (GCs).
In our numerical models, GCs form in
turbulent,
high-density  giant molecular clouds (GMCs),
which are 
embedded in a massive dark matter halo  
at high redshifts.
We show how high-density, compact stellar systems
are formed from  GMCs influenced both  by physical processes
associated with star formation and by tidal fields of their
host halos.
We also show that 
chemical pollution of GC-forming GMCs 
by stellar winds  from massive stars 
can result in star-to-star abundance inhomogeneities
among light elements (e.g., C, N, and O) of stars
in GCs.
The present model with a canonical initial mass function (IMF) 
also  shows a C-N anticorrelation that stars with smaller
[C/Fe] have larger [N/Fe] in a GC. 
Although these results imply that ``self-pollution''
of GC-forming GMCs by stellar winds from massive
stars can cause 
abundance inhomogeneities of GCs,
the present models with different
parameters  and canonical IMFs can not show N-rich stars with
${\rm [N/Fe]} \sim 0.8$ observed in some GCs (e.g.,
NGC 6752).
We discuss this apparent failure in the context of 
massive star formation preceding 
low-mass one within
GC-forming  GMCs (``bimodal star formation scenario'').
We also show that  although almost all stars ($\sim 97$\%)
show  normal He abundances (Y) of $\sim0.24$,
some stars later formed in GMCs 
can have Y as high as $\sim 0.3$ in some models.
The number fraction of He-rich stars with Y $>0.26$ 
is however found to be  small ($\sim 10^{-3}$) for most models.
We discuss this result in the context of
the possibly large  Y values
observed in a few Galactic GCs  (e.g.,  $\omega$ Cen and NGC 2808).
We also briefly discuss significantly larger 
degrees of abundance inhomogeneities 
of the models with stellar yield tables for rotating massive stars
in the bimodal star formation scenario.
\end{abstract}


\section{Introduction}

Dynamical and chemical properties of globular clusters
(GCs) in the Galaxy have been discussed in variously different
contexts, such as the Galaxy formation 
by accretion of low-mass dwarfs (e.g., Searle \& Zinn 1978),
self-pollution by  AGB  stars within proto-GC clouds 
(e.g., Cottrell \& Da Costa 1981),
and the formation process of GCs at high redshifts (e.g., Djorgovski 1993). 
The observed correlations and anticorrelations in
light elements (e.g.,  C, N, O, Na, Mg, and Al) of stars 
within GCs have long been considered to provide vital
clues for the early formation histories of GCs
(e.g., Gratton et al. 2004).
Recent observational studies, which  have revealed
star-to-star  abundance inhomogeneity among light elements
of stars
{\it on the main sequence} in the Galactic GCs,
have suggested that  the observed abundance inhomogeneity is due to
the second generation of
stars formed from ejecta of the first generation  of
stars (e.g., AGB and OB stars)  within GCs
(e.g., Da Costa et al. 2004):
proto-GC clouds could be chemically polluted 
by earlier generations of stars  
within forming GCs (which is often referred
to as ``self-pollution'').

It however remains unclear (1) how such ``self-pollution'' was possible
within proto-GC clouds (e.g., Smith \& Norris 1982)
and (2) whether massive stars or AGB ones are responsible
for the self-pollution processes 
(e.g., Prantzos \& Charbonnel 2006; Smith 2006).
Previous studies
discusses whether  chemical evolution models 
with {\it canonical IMFs} and  the self-pollution processes 
by AGB stars 
could  have some  problems in explaining the observed large
fraction of CN-strong stars
(Smith \& Norris 1982; Bekki 2006).
Decressin et al. (2006)
have recently suggested that self-pollution by 
stellar winds of massive stars
with masses of $20-120 {\rm M}_{\odot}$ 
and fast  rotation can possibly
explain the observed abundance patterns of GCs,
such as O-Na and Mg-Al anticorrelations.
It is however unclear how stellar winds from massive
stars influence chemical evolution of forming GCs
owing to the lack of extensive chemodynamical
simulations of forming GCs.

The purpose of this paper is thus to discuss, for the first time,
whether stellar winds from massive stars can be vital in chemical evolution
of forming GCs based on chemodynamical simulations of forming GCs
within GMCs.
We assume that GCs can be formed in GMCs within
low-mass galaxies embedded in dark matter halos
at high redshifts ($z>6$)
and thereby investigate the formation processes 
of GCs within the galaxies.
We  investigate how stellar winds from
massive stars, in particular, those with the masses
($m_{\rm s}$) larger than $8 {\rm M}_{\odot}$
influence chemical evolution of forming GCs
within GMCs. 
By comparing the observed abundance ratios of [N/Fe] and [C/Fe] 
and the corresponding simulation results,
we  raise important questions on IMFs and 
star formation histories of forming GCs. 
Since we focus on chemical abundances of GCs,
we do not intend to discuss structural properties 
and scaling relations (e.g., the Fundamental Plane)
of  GCs in the present paper.
Dynamical properties of GCs with different masses
will be discussed in our future papers (Bekki \& Chiba 2007, BC07,
in preparation).
We do not intend to discuss the roles of AGB stars
in the chemical evolution
of forming GCs in this paper (See Bekki 2006 and  Bekki et al. 2007
for the detailed discussions on this matter).

\section{The model}

Recent numerical simulations have suggested that
high-density gaseous regions within low-mass dwarf
galaxies embedded in dark matter halos at high redshifts ($z$) can be 
the formation sites of GCs 
(e.g., Bromm \& Clarke 2002).
Theoretical studies based on 
numerical simulations of GMCs with internal turbulent flows
suggested that 
bound star clusters  can  be formed
in GMCs with high star formation efficiencies 
(e.g., Klessen et al. 2000).
Guided by these previous studies,
we consider that GCs can be formed in turbulent GMCs 
within low-mass dwarfs at high $z$. 
Since the details of numerical methods and techniques
on chemodynamical simulations using GRAPE5-SPH codes
for GC formation 
are given in BC07,
we briefly summarize these in the following.

 GMCs with the sizes ($r_{\rm g}$) and
masses ($m_{\rm g}$) are assumed to have  homogeneous spherical
distributions for most models.
We set up the initial velocity fields due to turbulent flows
within  GMCs in the same way as Mac Low et al (1998) did.
We therefore assume that  
a turbulent velocity field  within a  GMC is a Gaussian random 
field with power spectrum $P(k) = P_0 k^{\alpha}$, where
$\alpha$ is set to be 2.0 for most models and
$P_0$ is a parameter controlling the total kinematical energy
due to the turbulent flow in the GMC. 
The virial ratio ($t_{\rm v}$) of $2T_{\rm k}/W$,
where $T_{\rm k}$ and $W$ are the total kinematical
energy and the absolute magnitude of the total potential
energy for a GMC, respectively, is  
determined by $P_0$ and set to be 0.5 in the present study.
Since an isothermal equation of state 
is suggested to be appropriate for star-forming
interstellar clouds of molecular gas  (e.g., Mac Low et al. 1998;
Klessen et al. 2000),
we adopt the equation with the initial temperature of 10 K.

A GMC  represented by $10^5$ SPH particles 
is assumed to be within a 
low-mass dark matter halo with a mass of
$10^9 {\rm M}_{\odot}$ and the universal ``NFW'' radial density
profile (Navarro et al. 1996) with the scale length ($r_{\rm s}$)
of 496 pc
and the concentration parameter ($c$)  of 10. 
A GMC is assumed to have a circular orbit and the initial distance
from the center of its host's  dark matter halo
is a free parameter described by $R_{\rm p}$.
Since  GMCs in a halo  are influenced strongly by tidal fields of the  halo,
the formation processes of GCs within GMCs can be different
in different locations within the halo.

 Gas particles are assumed to be converted into new stellar particles,
if (i) local dynamical time scales are shorter than local sound 
crossing time 
and (ii) local gas densities exceed a threshold gas
density (${\rho}_{\rm th}$) 
of star formation (e.g., Nakasato et al. 2000).
Recent high-resolution numerical simulations on the effects
of stellar winds from massive stars on interstellar
medium have demonstrated that about 30\% of the total 
energy ($\sim 10^{50}$ erg) from a massive star
can be converted into kinematical energy of gas around
the star (e.g., Freyer et al. 2003).
Guided by these simulations, we consider that
one massive star can give its surrounding gas
kinematical energy of  $3 \times  10^{49}$ erg
in the present study.
Stellar particles are assumed to lose gradually  their masses 
($m_{\rm s}$) owing to 
mass loss through stellar winds from  massive stars.
Since chemical abundances of stellar winds  from massive stars
are time-dependent and quite different in light elements
from those of gas from where the stars are born
(e.g., Schaller et al. 1992),
stars later formed from gas chemically polluted by
stellar winds from the first generation
of stars in a GMC can show abundances different  from those
in the first generation.

We use the numerical tables for time-dependent mass-loss rates
of non-rotating massive stars  and for the chemical yields of the stars
shown in Schaller et al. (1992) for a metallicity of $Z=0.001$
in order to calculate masses and abundances of stellar winds
from stellar particles at each time step
for a given IMF.
This is because these tables are only the ones that  show
explicitly the time evolution of abundances  
relevant to the present study 
(H, He, C, N, and O) for different massive stars.
Since chemical yield tables 
(including Na, Mg, and Al) for rotating massive stars
by Decressin et al. (2006) 
are only for those with a given mass ($m_{\rm I}=60 {\rm M}_{\odot}$)
and do not show the time-dependent evolution of mass-loss rates
and abundances of massive stars,
we do not use these tables in the present study. 
We adopt power-law IMFs ($\psi (m_{\rm I}) \propto  {m_{\rm I}}^{-s}$)
with the slopes of $s$ and
an upper- and lower-mass cut off ($m_{\rm u}$
and $m_{\rm l}$).
$m_{\rm l}$ is set  to be  $0.1 {\rm M}_{\odot}$
for all models whereas 
$m_{\rm u}$ is regarded as a free parameter controlling
self-pollution processes  by massive stars
in the present study.
Initial abundances of gas are set to be the same as those
used by Schaller et al. (1992) for consistency.
 
We focus mainly on the observed  C-N anticorrelation
(e.g., Smith \& Norris 1993) that stars with higher [N/Fe]
have smaller [C/Fe] in GCs,
firstly because these observations have been extensively
discussed in other theoretical papers (e.g., Bekki et al. 2007)
and secondly because [N/Fe] and  [C/Fe] 
can be investigated by the present models
based on stellar yield table by Schaller et al. (1992)
{\it which does not provide tables for Na, Mg, and Al}.
We will discuss O-Na and Mg-Al anticorrelations in 
our future papers, if the tables for {\it time-dependent
evolution of Na, Ma, and Al  for massive stars with variously
different masses and metallicities} are available:
for this reason, we do not use 
the latest important work  by Maeder \&  Meynet (2006)
for rotating and non-rotating massive stars.
It should be stressed here that
the observed C-N, O-Na, and Mg-Al anticorrelations
can be possibly  reproduced by chemical evolution
models using stellar yield tables for fast rotating
massive stars (Decressin et al. 2006).

Although we have investigated models with different parameter
values, we mainly show the ``standard model'' with
$m_{\rm g}=10^6 {\rm M}_{\odot}$, $r_{\rm g}=53$pc,
$s=2.35$, $m_{\rm u}=120 {\rm M}_{\odot}$, 
and $R_{\rm p}=0.2 r_{\rm s} = 99$pc.
This is mainly because this model clearly shows how 
a star-to-star abundance inhomogeneity within 
a forming GC is achieved during chemodynamical
evolution of the  GMC with star formation: 
the details of the  the parameter 
dependences of abundance patterns of simulated
GCs will be given in BC07.
The mass density of the initial GMC in the standard model
is much higher
than that adopted in the model
with $m_{\rm g}=5 \times 10^4 {\rm M}_{\odot}$ and $r_{\rm g}=50$pc
by Geyer \& Burkert (2002)
in which GC formation does not occur owing to low efficiencies
of star formation.
The mass densities ($n_{\rm g}$) of the preset models are however 
assumed to roughly follow the observed relation of 
$n_{\rm g} =3400 \times (\frac{r_{\rm g}}{1 {\rm pc}})^{-1.1}$ 
(atom cm$^{-3}$)
by Larson (1981).

We describe only five representative and important models 
(e.g., standard model) in this paper and the parameters values
and some important  results  in these models 
are given in table 1: $m_{\rm g}$ (column 1),
$R_{\rm p}$ (2), $s$ (3), $m_{\rm u}$ (4), $\delta$Y (5),
$\delta$[C/Fe] (6), and $\delta$[N/Fe] (7).
$\delta$Y ($\delta$[C/Fe]  and $\delta$[N/Fe])
is the  difference between the minimum and maximum  Y ([C/Fe]  and  [N/Fe]) 
at $T=11.0$ Myr.
For example,  the values shown in the table for the 
standard model mean that the maximum Y, [C/Fe],  and  [N/Fe] 
are 0.275, 0.168, and 0.324, respectively,  at  $T=11.0$ Myr.
[O/Fe] inhomogeneity is  not the main focus 
of this paper so that it is not so extensively discussed.
We however found that (i) $\delta$[O/Fe] is  0.030 for the standard
model and (ii) it is larger for models with larger $\delta$[C/Fe].

 We do not intend to discuss extensively the roles of
Type II supernovae (SNeII) in GC formation
in the present study, mainly
because we confirm that
SNeII are not so important for dynamical  evolution
and star formation in GMCs.
Previous simulations by Nakasato et al. (2000) 
also showed that self-enrichment of heavy elements
due to SNeII does not occur so that simulated
GCs show homogeneities in Fe-peak elements
(which is consistent with most GCs in the Galaxy). 
We find that if the time delay between 
star formation and the onset of supernovae explosion
is similar to a typical lifetime of massive stars
($\sim 10^7$ yr, which  corresponds to an average lifetime
for stars with $8 {\rm M}_{\odot} - 120 {\rm M}_{\odot}$
in models with $s=2.35$),
SNeII can be important only for expelling the remaining
gas of GC-forming GMCs into interstellar regions
of low-mass halos.
We thus show the results of models including feedback effects
of massive stars only for clarity:
the roles of SNeII in GC formation will be given
in BC07.

Thus, the present study adopts a fundamentally
different approach from previous one-zone chemical
evolution models (e.g., Bekki et al. 2007) in the sense that
not only  mixing of stellar ejecta (from massive stars) and the surrounding
gas but also star formation from the mixed gas 
are self-consistency
investigated through dynamical simulations with Jeans instability
for star formation, kinematical feedback effects of OB stars,
and non-instantaneous recycling of stellar ejecta.
The present simulations thus enable us to discuss whether 
the simulated GCs can explain {\it both} dynamical properties
and chemical ones, although the present models do not allow
us to discuss the observed  anti-correlations (e.g., C-N, O-Na, Mg-Al relations)
in a self-consistent manner.

\section{Results}

Figure 1 describes how the star formation rate
within a GMC evolves with time in the standard model.
First stars are formed within a GMC
at $T=2.4$ Myr when the turbulence
within the  GMC has decayed significantly
to form local high-density gaseous regions
(${\rho}_{\rm g} \ge 10^3$ atoms\,cm$^{-3}$). 
The star formation reaches   its maximum rate
($0.32 {\rm M}_{\odot}/{\rm yr}$)
at $T=5.0$ Myr and suddenly declines 
and finally stops completely till $T=8.0$ Myr 
owing to (i) feedback effects
of massive stars (i.e., the momentum input)
on the remaining gas and (ii) stripping
of the gas by the background tidal field of 
the dark matter halo. 
Almost 90\% of initial gas is converted into new stars
within a time scale of  $\sim 5$ Myr to form
a number of star clusters (SCs).

Figure 2 shows the final spatial distribution of
SCs at $T=43.5$ Myr  which corresponds to
one orbital rotation after 
most stars are formed within the GMC. 
The largest  SC  shows
a size of 10.6 pc,  an appreciably flattened shape,
two companion SCs,
and a total stellar
mass (including field stars and the companions) 
of $8.3 \times 10^5 {\rm M}_{\odot}$ within
50 pc from the SC's  center.
The two  small, low-mass, and bound SCs survived
from tidal destruction by their host halo may well
be identified as low-mass open clusters. 
Field stars that are diffusely distributed within
the halo were initially weakly bound or unbound  star-forming
complexes. 
These results imply  that not only a GC (and  open clusters)
but also field stars can be formed from a single high-density GMC
with a mass of $\sim 10^6 {\rm M}_{\odot}$.

Figure 3 shows that (i) the standard model 
can reproduce
a  C-N anticorrelation that stars with smaller [C/Fe]
have larger [N/Fe] in the simulated GC at $T=11.0$ Myr
and (ii) the star-to-star abundance spread in [C/Fe] and
[N/Fe] is about 0.03 and 0.22, respectively.
However it can not explain the existence of very N-rich stars
with [N/Fe]$\sim 0.8$ observed in NGC 6752 (Smith \& Norris 1993).
It should be  noted here that
the gaseous components show high [N/Fe] 
($\sim 0.8$) and low [C/Fe] ($\sim -0.2$).
These results imply that although chemical pollution
of gas  by massive stars earlier formed
does occur,  stars can not form efficiently from the
polluted gas later because the gas is  dispersed
into the interstellar space 
to have a lower density due both to feedback effects
of massive stars and to tidal stripping.

Figure 4 shows that
the simulated number fraction of N-rich ([N/Fe]$>0.2$) stars 
is small ($\sim  10^{-2}$).
This result is inconsistent with observations which
show the number fraction of CN-strong stars, which
are considered to be N-rich stars,
is as large as 0.5 in some GCs (e.g., NGC 6752; Smith \& Norris 1993).
This inconsistency results from the fact that
only a very small fraction of chemically polluted gas
is converted into new stars after the final burst of 
star formation ($T=6.0$ Myr).
It should be noted here that the number fraction
of N-rich components ([N/Fe] $>0.2$) in the gas is relatively high
($\sim 10^{-1}$).
These results imply that if a GC is formed from a GMC 
{\it which evolves in isolation
(i.e., without interacting with other GMCs in the halo)
and has a canonical IMF},
self-pollution by stellar winds of massive stars
can hardly explain the large fraction ($>0.1$)
of CN-strong (i.e., N-rich, C-depleted) stars
observed in some  Galactic GCs.
It should be however noted that the observed fraction of N-rich stars
is quite diverse (e.g., Norris 1988): some GCs with very low fraction
of N-rich stars  could  be consistent with some of 
the present  models.

Figure 4 shows that although stars 
show the He abundance spread,
the number fraction of moderately high He-rich (Y$>0.26$) stars 
is very small ($\sim 10^{-3}$): this model
is reasonable for GCs with ``normal'' Y.
Accordingly this  model  
can not reproduce 
the high fraction ($\sim 30$\%) of He-rich ($0.26 \le {\rm Y} \le 0.29$)
stars observed for NGC 2808 (D'Antona et al. 2005) 
and for NGC 6441 (Caloi \& D'Antona 2007),
which implies that models with canonical IMFs are not reasonable
for some GCs with high fraction of He-rich stars. 
Although Y of gaseous particles range from 0.24 to 0.57,
new stars can not form from He-rich gas with Y $>0.3$:
the origin of stars with possibly very high Y ($>0.3$)
in NGC 2808 (D'Antona et al. 2005)
can not be explained  by self-pollution processes 
of GC-forming GMCs  in the standard  model with a canonical IMF.
Although He-rich
gaseous components  can be finally formed,
the local densities of these components
become too low when they are formed
in the standard  model.
This is why the number fraction of He-rich
stars
is very small in this  model.
Figure 5  shows that abundance inhomogeneity in [O/Fe]
is much less significant in comparison with [N/Fe].
The simulated results
are not so consistent with the observations of
GCs stars, where a great reduction of the surface oxygen abundance is
observed in the most polluted objects.

Dependences of the results on model parameters are described
as follows. 
Although all four  models 
with different $m_{\rm g}$ and $R_{\rm p}$
and a canonical IMF with $s=2.35$ (i.e.,  the Salpeter IMF)
clearly show C-N anticorrelations
and star-to-star abundance spread in He, C, and N,
they can not show very N-rich stars with [N/Fe]$\sim 0.8$
observed for CN-strong AGB stars in NGC6752
(Smith \& Norris 1993):
$\delta$[N/Fe] ranges from 0.19 to 0.33 (corresponding
to $0.1\le {\rm [N/Fe]} \le 0.43$) in these models
(See the table 1).
The ``nucleus'' model shows a slightly larger degree of
abundance inhomogeneity
in the simulated GC, because gas ejected from massive
stars can not be brown away so effectively 
owing to the deep gravitational potential of the host halo:
the chemically polluted gas can be used for further star formation. 
The number fractions of stars with Y $>0.26$ 
(or with [N/Fe] $>0.2$) are at most $\sim 10^{-3}$  in these
models with $s=2.35$.

The ``top-heavy IMF'' model shows 
the presence of stars with   very high Y ($ \sim  0.5$),
because stellar winds
of stars with $m_{\rm I}=120 {\rm M}_{\odot}$
and thus with Y=0.72 in their later wind phases
can more effectively pollute the GC-forming GMC.
Figure 6  however shows that the fraction of Y-rich
stars with Y $>0.4$ is very small ($\sim 10^{-3}$). 
Figure 6 also shows that the number fraction of
N-rich stars with [N/Fe] $>0.2$ is 0.06,
which is more consistent with the observed
fraction of CN-strong stars in NGC 6752 
(Smith \& Norris 1993) than the standard model. 
We confirm that 
the fraction of N-rich stars can be as high as 0.4,
if we adopt the top-heavy  IMF models with initial isothermal
radial density profiles ($\rho(r) \propto r^{-2}$) of GMCs
for which stellar winds of massive stars can more effectively
pollute GC-forming GMCs in the very early stage of
star formation.  
Thus the present models with smaller $s$ can  better
explain stellar abundances of GCs with
larger degrees of abundance inhomogeneities
in Y, [C/Fe], and [N/Fe].

Although the above five models with $m_{\rm u}=120 {\rm M}_{\odot}$
clearly show abundance spread among stars within the simulated GCs,
the ``bottom-heavy IMF  model'' with $m_{\rm u} = 40  {\rm M}_{\odot}$ 
shows a very small degree of the abundance inhomogeneity;
$\delta$Y,  $\delta$[C/Fe], and $\delta$[N/Fe]
are all less than $10^{-5}$.
We confirm that (i) this result does not depend
on  $m_{\rm g}$ 
and (ii) $m_{\rm u} \ge 60  {\rm M}_{\odot}$ 
is required for the simulated GCs to show star-to-star abundance
inhomogeneities.   
These results imply that total numbers of massive stars
with $m_{\rm I}=60-120 {\rm M}_{\odot}$, which are determined
by $m_{\rm u}$ for a given $s$ (and $m_{\rm l}$),
are important determinants controlling whether
GCs can show observable star-to-star abundance
inhomogeneities.

\section{Discussions}

\subsection{Problems related to  the stellar wind scenario}

Recent observations have  reported that
star-to-star  abundance inhomogeneity can be 
seen in  less evolved stars
on the main sequence and subgiant-branch (e.g., Cannon et al. 1998
for 47 Tuc),
for which deep mixing is highly unlikely  to occur: 
the  self-pollution scenario might well be
more promising
than the mixing one.
The present numerical simulations have first demonstrated
that  star-to-star abundance inhomogeneity is possible
within {\it GC-forming GMCs} being polluted by stellar winds
of massive stars 
(From now on this scenario is referred to as
the stellar wind scenario just for convenience).

However
the simulated  fractions ($\sim 10^{-2}$) 
of N-rich ([N/Fe] $>0.2$) stars 
for the adopted  canonical IMF 
are  too small to be consistent with the observed
one for some GCs (e.g., NGC 6752; Smith \& Norris 1993).
This inconsistency implies either that
stellar populations other than massive stars (e.g. AGB
stars) could be responsible for the observed abundance
patterns of GCs
or that the present models did not include important
ingredients of GC formation processes
(and thus failed to explain the observations).
Since abundance inhomogeneity due to AGB stars were
already discussed in previous papers (Cottrell \& Da Costa 1981;
Bekki et al. 2007), 
we focus on the later implication below.

If IMFs of forming GCs are significantly
top-heavy (i.e., much smaller $s$),
the observed large fraction
of N-rich stars can be reproduced
(Smith \& Norris 1982; See also
Prantzos \& Charbonnel 2006 
in the context of
the observed O-Na anticorrelation).
Although our top-heavy IMF model with $s=1.35$ can show
a large fraction of N-rich stars,
the stellar wind scenario with a top-heavy IMF has serious problems 
as described below.
If we adopt the Salpeter IMF
with $m_{\rm l}$ =  $0.1 {\rm M}_{\odot}$
and $m_{\rm u}$ = $120 {\rm M}_{\odot}$, 
the mass fraction of massive stars with masses ranging from
$8 {\rm M}_{\odot}$ to $120 {\rm M}_{\odot}$
is about 0.84. Since the vast majority ($\sim 90$\%)
of gas of these massive 
stars can be expelled from GC-forming GMCs via 
SNe II (owing to very shallow gravitational potentials of GMCs),
the total masses of GCs can dramatically decrease
after their formation.
Previous theoretical works demonstrated that if self-gravitating
star clusters like GCs lose more than 50\% of their orignal masses,
they become soon disintegrated (e.g.,  Hills 1980).
Furthermore dynamical models of GCs with top-heavy IMFs
demonstrated that GCs losing a significant fraction of their
masses can be  disintegrated during their dynamical
evolution around  the Galaxy  (e.g., Chernoff \& Weinberg
1990). 
Thus we think that top-heavy IMFs in forming GCs  can not be plausible.

We can provide the following two  possible solutions for the 
above inconsistency in the stellar wind scenario. 
Firstly, most massive stars in GMCs are first formed 
before the major epoch of low-mass star formation so that
their stellar winds can chemically pollute 
a large fraction  of  remaining gas from which
low-mass stars are later formed.
This time delay between the major epoch
of massive star formation
and that of low-mass one was not modeled properly
in the present simulations
for which 
stellar particles (``stars'')
have almost identical masses.
In this second solution,
a large fraction of low-mass stars can be formed
from N-rich gas well polluted by massive stars earlier formed:
a larger degree of abundance inhomogeneity can be seen
in GCs.
It is however currently unclear in what physical conditions
of GMCs massive stars can be formed earlier than
low-mass ones: More quantitative discussions 
on this solution are given in \S 4.2 just for convenience.

Secondly, GMCs forming GCs were chemically polluted
not only by stars within the GMCs but also by those
outside the GMCs (e.g., field massive stars): ``external 
pollution'' by field massive stars
is required to explain the large fraction of N-rich stars.
Owing to the lack of extensive simulations on this
external pollution processes,
it is unclear whether and how this process can occur
in GMCs within low-mass galaxies at high redshifts.
Thus,
we can not determine  whether the first  or the 
second solution is more reasonable and realistic
until we perform new sets of simulations
on time delay between low- and high-mass star formation
and on the above  external pollution processes.

 Recent photometric observations of stars in  $\omega$ Cen
have discovered a double main sequence
(DMS) in the color magnitude diagrams (CMDs) of its stellar populations
(e.g., Bedin et al. 2004).
One of the most promising interpretation is that
stars on the bluer main sequence (bMS) of the DMS
represents a very helium-rich ($Y\ge0.3$) population
(e.g., Norris 2004).
Although the present chemodynamical simulations have first demonstrated
that the formation of stars with moderately high Y ($>0.26$)
is possible in GMCs chemically polluted by
stellar winds from  massive stars,
the derived number fraction
of He-rich stars ($Y> 0.26$) is too small ($\sim 10^{-3}$) 
to be consistent with the observed one (e.g., 0.3 for the bMS
in $\omega$ Cen).
Bekki \& Norris (2006)  
suggested that  $\omega$ Cen
was a nucleus of an ancient dwarf galaxy where 
He-rich gas could be efficiently transferred into its nucleus
and consequently used for star formation to form the bMS.

\subsection{Two-fold star formation: 
Massive stars first and low-mass ones second ?}

We have suggested above that if the vast majority of 
massive stars (``polluters''  of star-forming GMCs)
are formed well before the formation of low-mass ones,
the fraction of N-rich stars can be significantly
increased to be consistent with the observed one.
In order to discuss this point in a quantitative way,
we here assume that (i) a GC-forming GMC
has two major
epochs of star formation, (ii) the total gas mass consumed
for the formation of the first (second) generation of stars   
is $M_{1}$ ($M_{2}$),
(iii) the slope of the power-law IMF of the 1st (2nd)
generation is $s_1$ ($s_2$), (iv) the initial mass
fraction of N in the GMC 
is  $7.0 \times 10^{-5}$ (from the table by Schaller
et al. 1992 for $m_{\rm I}=120 {\rm M}_{\odot}$,  $Z=0.001$,
and $T=0$ yr),
and (v) the mass fraction of N
in the winds of stars in the 1st generation 
is  $7.7 \times 10^{-4}$ (from
the table by Schaller
et al. 1992 for $m_{\rm I}=120 {\rm M}_{\odot}$,  $Z=0.001$,
and $T=3.1 \times 10^6$ yr).
For a given $M_1$,
$s_1$, and  $s_2$ (fixed at 2.35),
we search for parameter values of $M_2$  for  which  $\delta$[N/Fe] 
between the 1st and the 2nd generations is as large as 0.8
observed for NGC 6752 (Smith \& Norris 1993).
In these models,  the vast majority of massive stars (``polluters'')
are formed in the first epoch of 
star formation owing to the adopted top-heavy IMF ($s_1<2.35$) 
whereas most low-mass stars are formed in the first and
the second epochs.

Figure  7 shows (i) the $s_1$ dependence of $M_2/M_1$ required
for  $\delta$[N/Fe]$\sim 0.8$ in ten models with different $s_1$
and  (ii) that of $N_2/N_1$, where
$N_1$ and $N_2$ are the number of low-mass stars 
with $m_{\rm I} = 0.1-8 {\rm M}_{\odot}$ 
in the first generation of stars 
and that in the 2nd,   respectively.
Figure 7 shows that (i) 
$M_1$ needs to be significantly larger than $M_2$
and (ii) the required $M_2/M_1$ is larger for
smaller $s_1$ (i.e., more top-heavy IMF).
These results imply that if stellar winds of
massive stars are responsible for the 
star-to-star abundance inhomogeneity
in a GC, 
most gas (96-99\%) of the  GC-forming GMC 
needs to be consumed for the 1st generation massive stars.
Figure 7 also shows that  $N_2/N_1$ can be as large as
1.0 (i.e., as observed in NGC 6752) for 
the models with very top-heavy IMFs of $s_1 < 1.2$.
This result implies that top-heavy IMFs in the first
epochs of massive star formation 
are essential for explaining
the observed large ($>1$) fraction of CN-strong stars
in some GCs.

One of possibly serious problems in this ``bimodal star
formation'' scenario is that the 2nd generation of 
stars in a GC-forming  GMC needs to be formed well before the supernovae
explosion of the 1st: otherwise the GC shows abundance spreads
in Fe-peak elements, which are not observed for typical GCs.
This problem would not be so serious {\it if stellar winds
from the 1st  can trigger star formation of the 2nd}.
It is however theoretically unclear in what physical conditions 
star formation of 
the 2nd can be triggered by stellar winds of the 1st owing
to the lack of numerical simulations on these processes.
Another problem is that a significant amount of mass
loss (more than 90\%) due to (i) top-heavy IMFs
and (ii) $M_1$ much larger than $M_2$ 
can lead rapid disintegration of forming GCs. 
We thus suggest that the proposed bimodal star formation 
scenario would not be so convincing without solving
the above disintegration problem due to  the assumed
top-heavy IMFs.

Then is the above IMF problem serious also for models in which
chemical yield tables for rotating massive stars are used ?
In order to answer this question,
we investigate 
dependences of  $N_2/N_1$ and  $M_2/M_1$ on $s_1$
by using (i) final mass fraction
of stellar winds for rotating massive stars with different
masses and (ii) N abundances for these stars shown 
in Decressin et al. (2006). 
Figure 8 shows that for slightly  flatter IMFs
(i.e., $s_1 \ge 1.85$),   $N_2/N_1$ can be as large as $\sim 1$
(e.g.,   $N_2/N_1=1.2$ for  $s_1 = 1.85$)
and the required $M_2/M_1$ is significantly larger ($>0.3$),
which suggests that the above disintegration problem  is much less
serious in the bimodal star formation scenario with
stellar yields from rotating massive stars. 
We thus suggest that future chemodynamical models based on 
the bimodal star formation scenario are 
promising in terms of  explaining 
the observed spread in N abundances of GCs.

\subsection{Relations to the Galactic open clusters and halo
field stars}

We have shown that
the  models with $m_{\rm u} = 40  {\rm M}_{\odot}$ 
show a very small degree of the abundance inhomogeneity;
$\delta$Y,  $\delta$[C/Fe], and $\delta$[N/Fe]
are all less than $10^{-5}$.
These results imply that (i) it depends on
$m_{\rm u}$ of GC-forming GMCs whether the GCs show 
abundance inhomogeneities and (ii) the observed
differences in the degree of abundance inhomogeneities
in the Galactic GCs can be due also to the differences
in the number fraction of very massive stars 
formed in GC-forming GMCs.
The fraction of N-rich stars in a GC can be observationally
quantified by  an  ``$r$'' parameter by counting
the number fraction of  CN-strong stars
vs CN-weak stars  (Norris 1988),
and the values of these ``$r$'' parameters are observed to
be different between the Galactic GCs (Norris 1988).
We suggest that the observed difference of $r$
is due to the difference in $m_{\rm u}$ and
the number fraction of very massive stars between GC-forming
GMCs.

Although the Galactic GCs show abundance inhomogeneities,
the old Galactic open clusters (OCs)
do not show
significant abundance inhomogeneities
(e.g., De Silva et al. 2006).
One of the possible explanations  for OCs without abundance inhomogeneities  
is that 
massive stars with $m_{\rm I}=60-120 {\rm M}_{\odot}$
could not be formed in GMCs of OCs for some physical reasons.
For the  Salpeter IMF,
the total number of massive stars with $m_{\rm I}=60-120 {\rm M}_{\odot}$
is just 0.3 (i.e., less than 1) for low-mass OCs 
with masses ($M_{\rm oc}$) of $1000 {\rm M}_{\odot}$.
This means that
the probability of low-mass OCs having very massive stars with 
$m_{\rm I}=60-120 {\rm M}_{\odot}$
is very low (owing to small-number statistics).
We thus suggest that the origin of the observed lack of
abundance inhomogeneities in the Galactic old OCs can be understood
in term of incapability of their host GMCs to form 
massive stars with $m_{\rm I}=60-120 {\rm M}_{\odot}$.

The vast majority (90\%) of stars are observed to form 
in SCs embedded within GMCs (Lada \& Lada 2003):
strongly bound SCs can finally evolve into GCs or OCs
whereas weakly bound, low-mass  ones can be disintegrated into
field stars in galaxies.
Previous simulations (Bekki \& Chiba 2000, 2001) showed that
the Galactic stellar halo can be formed from
by both dissipative and dissipationless merging of subgalactic 
clumps and their resultant tidal disruption 
in the course of gravitational contraction 
of the Galaxy at high redshift ($z>1$): the halo field stars
were initially field stars originating from low-mass, unbound
(or weakly bound)  star clusters formed from GMCs  
not polluted by massive stars 
in high-$z$ building blocks of the Galaxy.
These previous results combined with
the present ones therefore imply
that there can be significant differences
in abundance patterns of light elements between
the Galactic halo field stars and stars within
the GCs.

Observational studies have extensively discussed similarities and  differences
in abundances between the Galactic halo field  stars
and the GCs (e.g., Frogel 1993; Sneden 2005). 
For example, Suntzeff (1993) showed that
(i) the Galactic halo field  stars show  an apparently no  clear
O-Na anticorrelation, 
(ii) oxygen abundance is relatively constant at [O/Fe] $\sim 0.5$
for the halo field  stars with [Fe/H] $<-1$,
and (iii) these two properties of the halo field stars
are thus in a  striking contrast with
those of the GCs.
We suggest that these differences reflect the fact that
only GCs originate from strongly bound, massive SCs formed
in GMCs polluted by stellar winds of very massive stars 
in low-mass subhalos:
although both the halo field  stars and the GCs were
initially ``clusters''
within low-mass building blocks of the Galaxy at high-$z$
and later stripped to become the halo populations,
the differences
in physical properties of their parent SCs/GMCs 
result in the differences of abundances between the two
halo populations.

\section{Conclusions}

We have first investigated 
the influences of  stellar winds of  
massive  stars on early chemical evolution of GC-forming
GMCs based on GRAPE SPH chemodynamical simulations.
In our numerical models, GCs form in
turbulent,
high-density  giant molecular clouds (GMCs),
which are 
embedded in a massive dark matter halo  
at high redshifts.
We have focused on the chemical evolution of GC-forming
GMCs in the present study.
We summarise our principal results
of the models as follows.

(1) Chemical pollution of GC-forming GMCs 
by stellar winds  from massive stars 
can result in star-to-star abundance inhomogeneities
among light elements (e.g., C, N, and O) of stars
in GCs.
The present model with a canonical IMF 
 ($s=2.35$) 
also  shows a C-N anticorrelation that stars with smaller
[C/Fe] have larger [N/Fe] in a GC. 
These results imply that ``self-pollution''
of GC-forming GMCs by stellar winds from massive
stars can cause
abundance inhomogeneities of GCs.

(2) The present models with different
parameters  and  canonical IMFs  ($s=2.35$) 
however can not show N-rich stars with
${\rm [N/Fe]} \sim 0.8$ observed in some Galactic GCs
(e.g. NGC 6752).
The simulated number fraction of N-rich ([N/Fe]$>0.2$) stars 
is small ($\sim  10^{-2}$) in most models with $s=2.35$.
The clear bimodality in [N/Fe] can not be seen in the models,
mainly because the stellar winds of the 
first generation of stars can not
trigger a  strong burst of star formation for the 2nd generation 
in the present models with a  single,  structureless GMC. 
Models with top-heavy IMFs ($s=1.35$)
show a significantly larger fraction ($\sim 0.1$) of N-rich stars.
Although models with top-heavy IMFs
are more consistent with observations,
GCs in these models
are suggested to be disintegrated soon after GC formation.

(3) Although almost all stars ($\sim 97$\%)
show  normal He abundances (Y) of $\sim0.24$,
some stars later formed in GMCs 
can have Y as high as $\sim 0.3$ in some models.
The number fraction of He-rich stars with Y $>0.26$ 
is however found to be  small ($\sim 10^{-3}$) for most models.
These results imply that   (i)
the Galactic GCs 
can have  Y abundance inhomogeneities 
and (ii) the degree of inhomogeneities can depend
on IMFs and $m_{\rm u}$ of GC-forming GMCs.

(4) The present models with canonical IMFs  can hardly explain
the observed large fraction of CN-strong stars
in NGC 6752 and  the large fraction of He-rich 
stars in NGC 2808. We therefore 
discussed these apparent failures in the context of 
massive star formation preceding
low-mass one within
GC-forming  GMCs (i.e., the bimodal star formation scenario)
and the new external pollution scenario. 
The formation of the 2nd generations of stars  needs to be 
triggered by the stellar winds of the  1st ones to explain the observed
very small spread in [Fe/H] in the bimodal star formation
scenario. 
The above IMF problem is however demonstrated to be much less
severe in the bimodal star formation scenario with
stellar yield tables for rotating massive stars.

(5) The origin of the observed differences in abundances patterns
of light elements (e.g., C, N, and O) between
the Galactic halo field  stars and the GCs can reflect the fact that 
the field stars originate from low-mass, unbound star clusters formed
in GMCs not polluted by stellar winds of very massive stars
with $m_{\rm I} \ge 60 {\rm M}_{\odot}$.
Although both the halo field  stars and the GCs were initially in 
the low-mass subhalos and later stripped to become the halo populations,
only GCs originate from strongly bound, more massive
star clusters formed within GMCs polluted efficiently by
stellar winds of very massive stars.

Thus, the present chemodynamical simulations, which first
have investigated self-consistently star formation histories
and time evolution of He, C, N, and O within GC-forming GMCs,
have shown that the observed abundance 
inhomogeneities of GCs stars cannot be explained on the basis of the
stellar winds by massive stars, 
if a realistic IMF and the yields by
Schaller et al (1992) are used.
In the present paper, we have not considered  the self-pollution
of GC-forming gas clouds by ejecta of AGB stars,  for which
top-heavy IMFs are required to explain the observed large fraction
of N-rich stars in some GCs (e.g., Smith \&  Norris; Bekki et al. 2007).
If a top-heavy IMF is the only way to explain 
the observed degrees of abundance inhomogeneities of GCs
both in the stellar wind scenario
and in the AGB one and if GCs with top-heavy IMFs inevitably
disintegrate soon after their formation,
the external pollution scenario discussed briefly
in this paper needs to be investigated by chemodynamical simulations 
in a more quantitative way both for the two scenarios.
We have also suggested that 
different IMFs in two major epochs of star formation for
GC-forming GMCs
need to be explored in future chemodynamical models, in particular,
with  chemical yield tables for rotating massive stars.

\acknowledgments
We are  grateful to the anonymous referee for valuable comments,
which contribute to improve the present paper.
K.B. acknowledges the financial support of the Australian Research
Council throughout the course of this work. 
The numerical simulations reported here were carried out on GRAPE
systems kindly made available by the Astronomical Data Analysis
Center (ADAC) at National Astronomical Observatory of Japan (NAOJ).


\begin{deluxetable}{ccccccccc}
\tabletypesize{\footnotesize}
\tablecaption{Model parameters and results \label{tbl-1}}
\tablewidth{0pt}
\tablehead{
\colhead{model} & 
\colhead{$m_{\rm g}$ ($\times 10^6 {\rm M}_{\odot}$)} & 
\colhead{$R_{\rm p}$ (pc)} &
\colhead{$s$} &
\colhead{$m_{\rm u}$} &
\colhead{$\delta$Y}  &
\colhead{$\delta$[C/Fe]} & 
\colhead{$\delta$[N/Fe]} }
\startdata
standard  & 1.0 &  99 & 2.35 & 120 & 0.032  & 0.028  & 0.223 \\
lower mass  & 0.1 &  99 & 2.35 & 120 & 0.026 & 0.023 & 0.189 \\
lowest mass  & 0.05 &  99 & 2.35 & 120 & 0.032  & 0.028  & 0.222 \\
nucleus  & 1.0 &  0 & 2.35 & 120 & 0.054   & 0.049  &0.330 \\
top-heavy IMF  & 1.0 &  99 & 1.35 & 120  & 0.270 &0.332 & 0.827 \\
bottom-heavy IMF  & 1.0 &  99 & 2.35 & 40  & 
$<10^{-5}$ & $<10^{-5}$ & $<10^{-5}$ \\
\enddata
\end{deluxetable}


\begin{figure}
\plotone{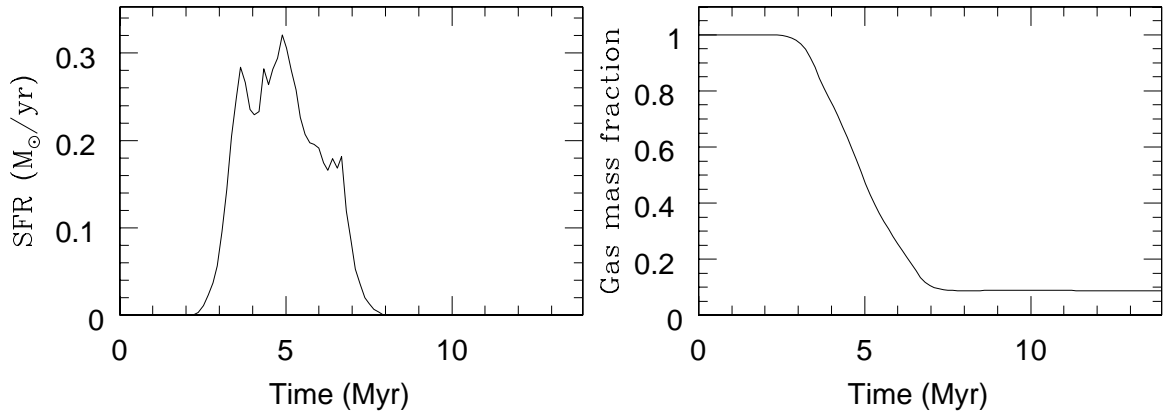} 
\caption{
The time evolution of star formation rate (left)
and gas mass fraction (right) of a GMC
in the standard model.
\label{fig-1}}
\end{figure}


\begin{figure}
\plotone{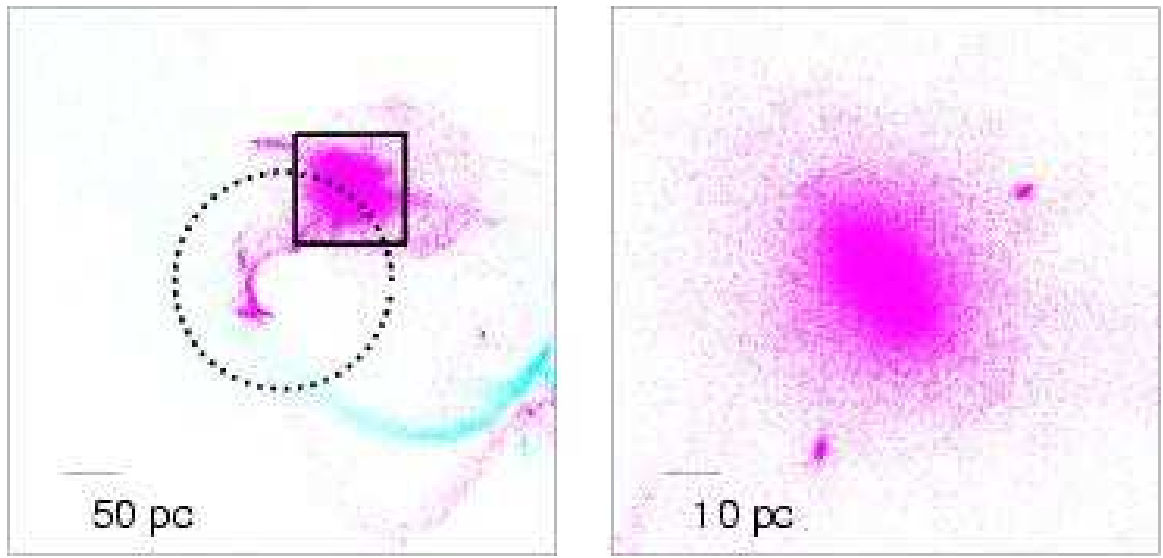} 
\caption{
Final distributions of stars (magenta) and gas (cyan)
at $T=43.5$ Myr
projected onto the initial orbital plane of the GMC within 
the halo in the standard model. 
The center of the frame in the left panel is coincident
with the center of the halo, and the thick dotted line 
represent the orbit of the GMC.
The center of the frame in the right panel
is coincident with the mass center of the most
massive SC (star cluster)
shown within a square marked by a thick solid
line in the left panel.
\label{fig-2}}
\end{figure}


\begin{figure}
\plotone{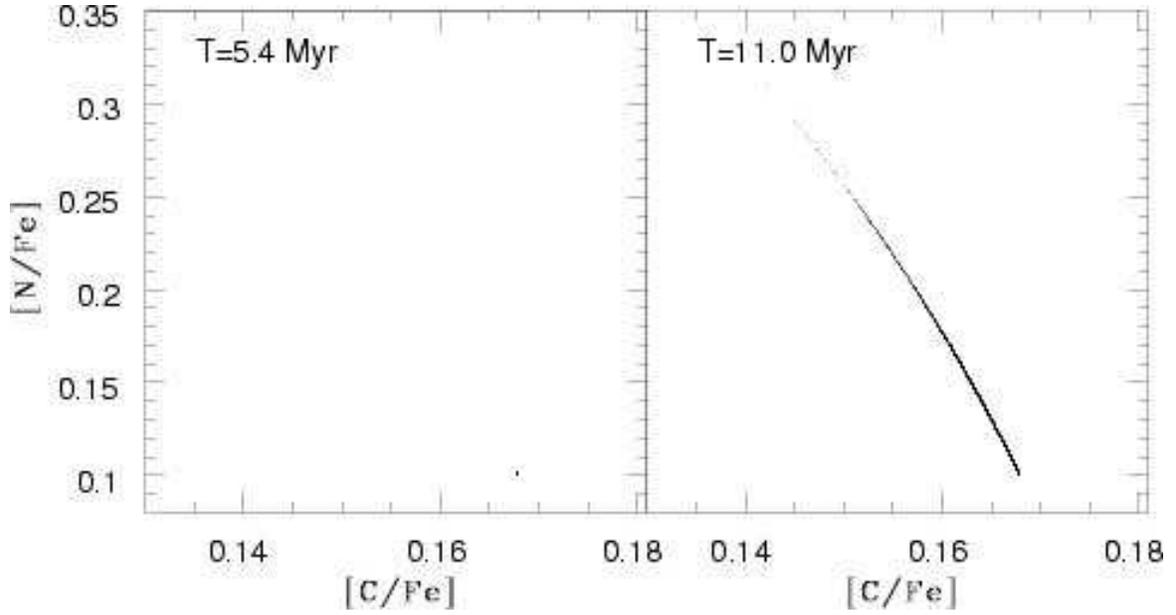} 
\caption{
The distributions of stars 
on the [C/Fe]-[N/Fe] plane at $T=5.4$ Myr (left)
and $T=11.0$ Myr (right) in the standard model.
\label{fig-3}}
\end{figure}


\begin{figure}
\plotone{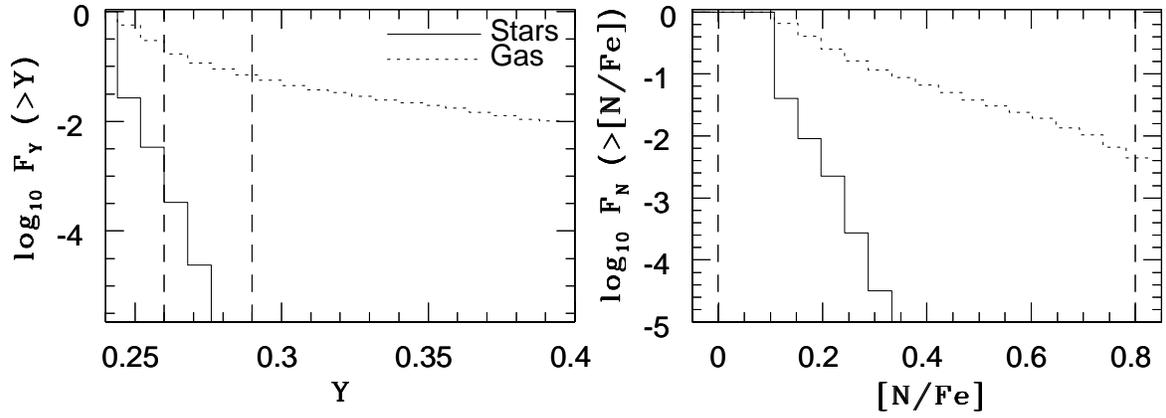} 
\caption{
Normalized cumulative number distribution (in log scale) of
He ($F_{\rm Y} {\rm (>Y)}$; left)
and [N/Fe] ($F_{\rm N} {\rm (> [N/Fe])}$; right) 
for stars (solid)
and gas (dotted) in the standard model.
For comparison, 
the observed range of $0.26 < {\rm Y} < 0.29$
for He-rich stars consisting of about 30\% of
the entire stellar population in NGC 2808 (D'Antona et al. 2005)
are shown by two dashed lines in the left panel.
The observed range of $0.0 < {\rm [N/Fe]} < 0.8$
for AGB stars
in  NGC 6752 (Smith \& Norris 1993) is shown by two dashed lines
in the right panel.
Note that the number fraction of He-rich stars with Y$>0.26$ is
very small ($\sim 10^{-3}$).
\label{fig-4}}
\end{figure}


\begin{figure}
\epsscale{.50}
\plotone{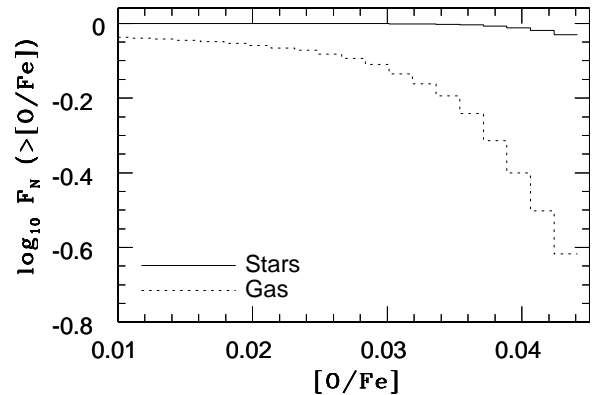} 
\caption{
The same as Fig. 4 but for [O/Fe]. 
\label{fig-5}}
\epsscale{1.0}
\end{figure}


\begin{figure}
\plotone{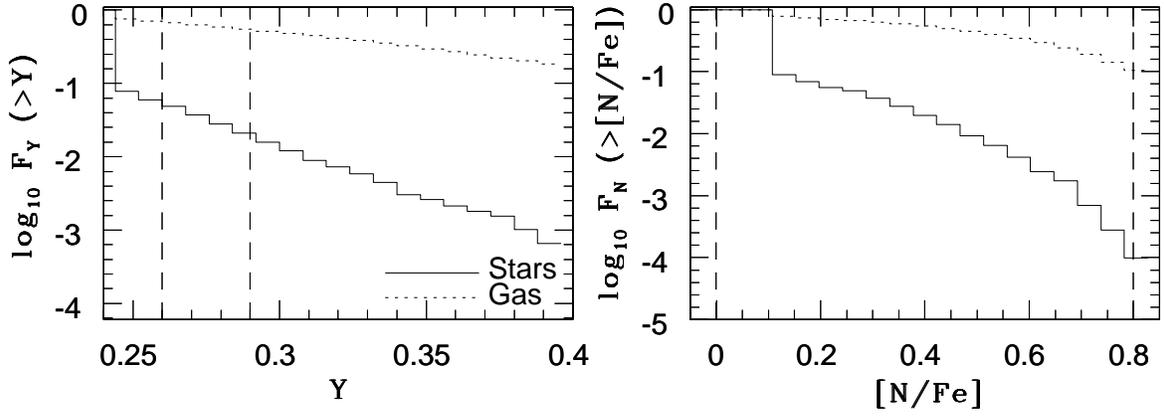} 
\caption{
The same as Fig. 4 but for the top-heavy IMF model with
$s=1.35$.
\label{fig-6}}
\end{figure}


\begin{figure}
\plotone{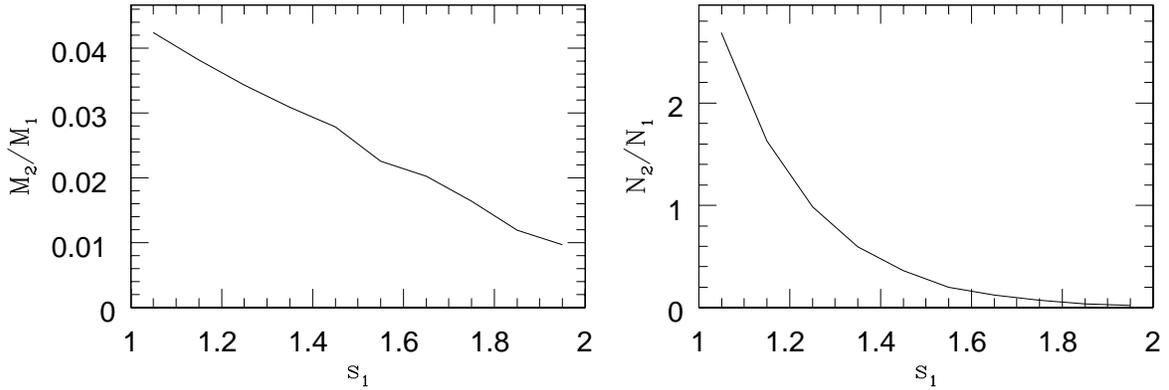} 
\caption{
Dependences of $M_2/M_1$ (left) and $N_2/N_1$ (right)
on $s_1$, where $M_1$, $M_2$,  $s_1$, $N_1$, and 
$N_2$ are the initial 
total gas mass of the 1st generation of stars, that of the 2nd generation,
the slope of the IMF in the 1st, total number of low-mass stars
in the 1st, and that in the 2nd, respectively.
The IMF slope for the 2nd
generation ($s_2$) is assumed to be 2.35.
\label{fig-7}}
\end{figure}


\begin{figure}
\plotone{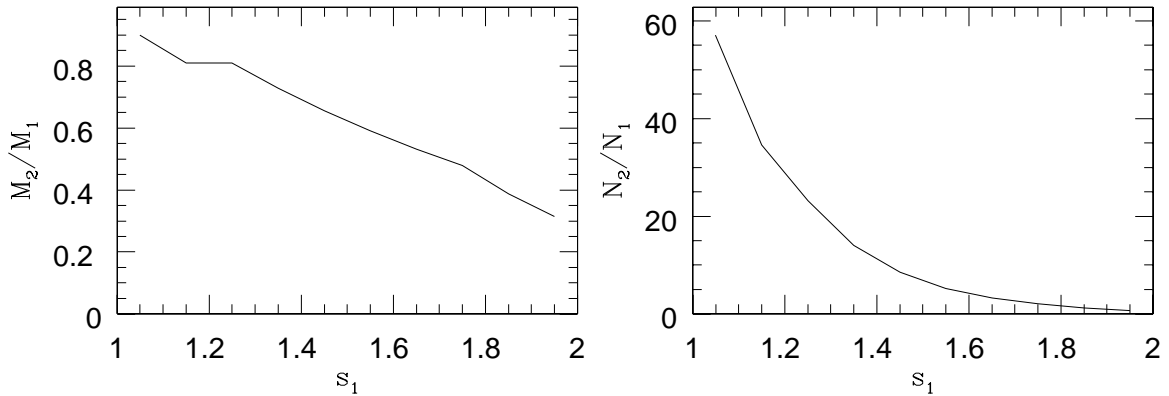} 
\caption{
The same as Figure 7 but for models in which chemical yield
tables for rotating massive stars (Decressin et al. 2006)
are used.
\label{fig-8}}
\end{figure}

\end{document}